\documentclass[12pt]{iopart}
\usepackage[latin1]{inputenc}
\usepackage[T1]{fontenc}
\usepackage[english]{babel}
\begin{document}
\title[Isotropisation of Bianchi class $A$ models with curvature...]{Isotropisation of Bianchi class $A$ models with curvature for a minimally coupled scalar tensor theory}
\author{Stéphane Fay\\
Laboratoire Univers et Théories, CNRS-FRE 2462\\
Observatoire de Paris, F-92195 Meudon Cedex\\
France\\
\small{Steph.Fay@Wanadoo.fr}}
\begin{abstract}
We look for necessary isotropisation conditions of Bianchi class $A$ models with curvature in presence of a massive and minimally coupled scalar field when a function $\ell$ of the scalar field tends to a constant, diverges monotonically or with sufficiently small oscillations. Isotropisation leads the metric functions to tend to a power or exponential law of the proper time $t$ and the potential respectively to vanish as $t^{-2}$ or to a constant. Moreover, isotropisation always requires late time accelerated expansion and flatness of the Universe.
\\
\\
\\
Published in Classical and Quantum Gravity copyright 2003 IOP Publishing Ltd\\
Classical and Quantum Gravity, Vol 20, 7, 2003.\\
http://www.iop.org
\end{abstract}
\maketitle
%------------------------------------------------------------------------------------------------------------------------------------------------------------------------------------------%
\section{Introduction}\label{s0} In this paper we study isotropisation of Bianchi class A models with curvature when a minimally coupled and massive scalar field $\phi$ is considered.

Locally, General Relativity (GR) with a perfect fluid seems a good description of our Universe. At cosmological scale, accelerated dynamical expansion is observed\cite{Per99, Rie98} and additional fields are required to explain it. Among them, a scalar field seems a good alternative although it is not the only one: higher order theories\cite{CotMir97, Fed00} or dissipative fluid\cite{ChiJakPav00, SenSen00} also give birth to inflationary behaviour. Scalar fields are required by standard model for elementary particles as well as by unification theories for which, for instance, compactification schemes\cite{EllKalOliYok99, Wan94} are considered. These last theories also give a natural order of magnitude for the cosmological constant\cite{WesLiu01} at early times which may be 55 to 120 orders of magnitude bigger than its present observed value: this is the so-called cosmological constant problem. A solution is to consider that this "constant" varies across the Universe history. A massive scalar field is then an interesting possibility to simulate such a mechanism. All these elements show the interest of a minimally coupled scalar-tensor theory with a Brans-Dicke coupling function $\omega$ and a potential $U$ depending on the scalar field $\phi$.

What about the geometrical framework of this paper? FLRW models geometrically describe the observed homogeneity and isotropy of our Universe. However they are very special ones among the set of all possible models and do not allow to explain the observed large-scale structures. Moreover, at early times, before the decoupling between matter and radiation, we have no indication about Universe's geometry. Was it as so symmetric as the FLRW models imply? Thus, it seems interesting to generalise them by only keeping their spatial homogeneity property. Bianchi models describe anisotropic cosmological models and may allow to understand the process leading to an isotropic Universe. The most studied Bianchi models are those containing the FLRW solutions\cite{ColHaw73}, i.e. the types $I$, $V$, $VII_{0,h}$ and $IX$. We have examined the Bianchi type $I$ model in \cite{Fay01}. Here, we will be interested in the Bianchi class $A$ models with curvature.

Our goal is to look for necessary conditions allowing the isotropisation of Bianchi class $A$ models with curvature when a minimally coupled and massive scalar tensor theory is considered. We will then deduce the common asymptotical behaviour of the metric functions when isotropisation is reached and compare our results with those obtained for the Bianchi type $I$ model\cite{Fay01}. From a technical point of view, we will use the methods of \cite{Fay01}: we will get the field equations from ADM Hamiltonian formalism\cite{Nar72, MatRyaTot73} and rewrite them with a new set of variables. Then we will look for equilibrium points corresponding to isotropic stable states.

A large amount of work has been done on equilibrium states of Bianchi models. Wainwright, Ellis and collaborators have studied equilibrium points of homogeneous models for General Relativity with perfect fluid, tilted or not and found asymptotically isotropic solutions. A good summary of their work is \cite{WaiEll97}. They use Hubble-normalized variables to study the dynamics of Einstein field equations. The normalisation factor is the Hubble parameter and the equation allowing to show that variables are normalized is the generalized Friedman equation. In this paper we will also consider normalized variables but we will use the Hamiltonian as normalisation factor. The expression for Hamiltonian will be the constraint from which we deduce that variables are normalized. More recently Barrow and Kodama\cite{BarKod01, BarKod00} have examined the influence of topology on isotropy and flatness of the Universe. They have shown that "the topology of the Universe can impose significant restrictions upon the type of anisotropies it can sustain". We will not consider topology in this work but these results are really interesting from the point of view of relations between dynamics and topology which has also been examined by Ashtekar and Samuel\cite{AshSam91}.

The plane of the paper is the following. The second section will be parsed into three subsections. The first one will be devoted to the Bianchi type $II$ model, the second one to the Bianchi types $VI_0$ and $VII_0$ and the third one to the Bianchi type $VIII$ and $IX$ models. Each of these subsections will be divided into two subsections devoted to the field equations and the study of the equilibrium points. We will discuss the physical meaning of our results in section \ref{s3}.
%------------------------------------------------------------------------------------------------------------------------------------------------------------------------------------------%
\section{Mathematical study of isotropisation for class $A$ Bianchi models} \label{s1}
We begin calculating the Hamiltonian field equations. The Lagrangian of the minimally coupled scalar-tensor theory is given by:
%----------------------------------EQUATION----------------------------------------%
\begin{equation} \label{action1}
S=(16\pi)^{-1}\int \left[R-(3/2+\omega(\phi))\phi^{,\mu}\phi_{,\mu}\phi^{-2} -U(\phi)\right]\sqrt{-g}d^4 x
\end{equation}
Although it may be more natural to redefine $\phi$ so that the kinetic term takes a standard form $\phi^{,\mu}\phi_{,\mu}$, we prefer considering an unspecified Brans-Dicke coupling function such that our results be valid for any form of $\omega(\phi)$ even when it is analytically impossible to get $\phi(\omega)$. The general form of the metric for Bianchi models is written:
%----------------------------------EQUATION----------------------------------------%
\begin{equation}\label{metrique}
ds^2 = -(N^2 -N_i N^i )d\Omega^2 + 2N_i d\Omega\omega^i + R_0 ^2 g_{ij}\omega^i \omega^j 
\end{equation}
The $\omega_i$ are the 1-forms defining each Bianchi model. $N$ and $N^i$ are respectively the lapse and shift functions. To find the ADM Hamiltonian corresponding to the action (\ref{action1}), we proceed as in \cite{Nar72} and \cite{MatRyaTot73}. We rewrite the action as follows:
%-----------------------EQUATION-------------------------------------%
\begin{equation}\label{action2}
S=(16\pi)^{-1}\int(\Pi^{ij}\frac{\partial{g_{ij}}}{\partial{t}}+\Pi^{\phi}\frac{\partial{\phi}}{\partial{t}}-NC^0-N_iC^i)d^4x
\end{equation}
The $\Pi_{ij}$ and $\Pi_\phi$ are respectively the conjugate momenta of the metric functions and scalar field. The lapse and shift functions now play the role of Lagrange multipliers. By varying (\ref{action2}) with respect to $N$ and $N_i$, we get the constraints $C^0=0$ and $C^i=0$ with:
%----------------------------------EQUATION----------------------------------------%
\begin{equation}\label{C0}
C^0 =-\sqrt{^{(3)}g}^{(3)}R-\frac{1}{\sqrt{^{(3)}g}}(\frac{1}{2}(\Pi^k _k )^2 -\Pi^{ij}\Pi_{ij})+\frac{1}{\sqrt{^{(3)}g}}\frac{\Pi_\phi ^2 \phi^2 }{6+4\omega}+\sqrt{^{(3)}g}U(\phi)
\end{equation}
%----------------------------------EQUATION----------------------------------------%
\begin{equation}
C^i =\Pi^{ij}_{\mid j}
\end{equation}
We rewrite the metric functions $g_{ij}$ as $e^{-2\Omega+2\beta_{ij}}$. It means that $\Omega$ stands for the isotropic part of the metric and $\beta_{ij}$ for the anisotropic parts. Then, using Misner parameterisation \cite{Mis62}:
%-----------------------EQUATION-------------------------------------%
\begin{equation}
p_k^i=2\pi\Pi_k^i-2/3\pi\delta_k^i\Pi_l^l
\end{equation}
%-----------------------EQUATION-------------------------------------%
\begin{equation}
6p_{ij}=diag(p_++\sqrt{3}p_-,p_+-\sqrt{3}p_-,-2p_+)
\end{equation}
%-----------------------EQUATION-------------------------------------%
\begin{equation}
\beta_{ij}=diag(\beta_++\sqrt{3}\beta_-,\beta_+-\sqrt{3}\beta_-,-2\beta_+)
\end{equation}
and rewriting the Hamiltonian as $H=2\pi\Pi_k^k$, from the expression (\ref{C0}) and the constraint $C^0=0$, we get:
%----------------------------------EQUATION----------------------------------------%
\begin{equation} \label{hamiltonien}
H^2 = p_+ ^2 +p_- ^2 +12\frac{p_\phi ^2 \phi^2}{3+2\omega}+24\pi^2 R_0 ^6 e^{-6\Omega}U+V(\Omega,\beta_+,\beta_-)
\end{equation}
with $p_\phi=\pi \Pi_\phi$. The form of $V(\Omega,\beta_+,\beta_-)$ specifies each Bianchi model and is given in table 1. From (\ref{hamiltonien}), we derive the Hamiltonian equations:
%----------------------------------EQUATION----------------------------------------%
\begin{equation} \label{betapm}
\dot{\beta}_ \pm = \frac{\partial H}{\partial p_ \pm}=\frac{p_\pm}{H}
\end{equation}
%----------------------------------EQUATION----------------------------------------%
\begin{equation} \label{phip}
\dot{\phi}=\frac{\partial H}{\partial p_\phi}=\frac{12\phi^2 p_\phi }{(3+2\omega)H}
\end{equation}
%----------------------------------EQUATION----------------------------------------%
\begin{equation} \label{ppp}
\dot{p}_+=-\frac{\partial H}{\partial \beta_ \pm}=-\frac{\partial V(\Omega,\beta_+,\beta_-)/\partial \beta_+}{2H}
\end{equation}
%----------------------------------EQUATION----------------------------------------%
\begin{equation} \label{pmp}
\dot{p}_-=-\frac{\partial H}{\partial \beta_ \pm}=-\frac{\partial V(\Omega,\beta_+,\beta_-)/\partial \beta_-}{2H}
\end{equation}
%----------------------------------EQUATION----------------------------------------%
\begin{equation} \label{pphip}
\dot{p}_\phi=-\frac{\partial H}{\partial \phi}=-12\frac{\phi p_\phi ^2}{(3+2\omega)H}+12\frac{\omega_\phi \phi^2 p_\phi ^2 }{(3+2\omega)^2 H}-12\pi^2 R_0 ^6 \frac{e^{-6\Omega}U_\phi }{H}
\end{equation}
%----------------------------------EQUATION----------------------------------------%
\begin{equation} \label{hamp}
\dot{H}=\frac{dH}{d\Omega}=\frac{\partial H}{\partial \Omega}=-72\pi^2 R_0 ^6 \frac{e^{-6\Omega}U}{H}+\frac{\partial V(\Omega,\beta_+,\beta_-)/\partial \Omega}{2H}
\end{equation}
We set $N_i=0$ and calculate that, whatever the Bianchi model, the lapse function is given by:
%-------------------------EQUATION-------------------------------------%
\begin{equation} \label{lapse}
N=\frac{12\pi R_0^3e^{-3\Omega}}{H}
\end{equation}
Then, the relation between the time $\Omega$ and the proper time $t$ is $dt=-Nd\Omega$.\\
Before starting the analysis of each Bianchi model, let us talk about some necessary conditions for isotropisation. By definition, the Universe isotropises if each metric function tends toward a common form, let us say $R^2$. From Misner parameterisation, we deduce that it implies $e^{-\Omega}\rightarrow R$ and $\beta_\pm\rightarrow 0$. A convenient measure of anisotropy is given by the a mesure of the root mean square anisotropy\cite{RyaShe75} $d\beta_+/dt^2+d\beta_-/dt^2=(d\beta_+/d\Omega^2+d\beta_-/d\Omega^2)(d\Omega/dt)^2$ which have to decay such that isotropy occurs. Assuming that isotropisation is an asymptotic phenomenon arising when proper time diverges, it means that asymptotically $d\beta_\pm/dt\rightarrow 0$. It is this last necessary condition only that we will use in this work. It is not sufficient for isotropisation since it does not prevent $\beta_\pm$ to diverge or to tend toward a big constant but it is necessary if we want that the Hubble factors $H_i=\frac{dg_{ii}}{dt} g_{ii}^{-1}$ be asymptotically the same, i.e. tend toward the same value $d\Omega/dt$ in accordance with the fact that the Hubble constant is the same in any direction. Since the equations (\ref{betapm}) and the expression for $N$ lead to:
%-------------------------EQUATION-------------------------------------%
\begin{equation} \label{isoC1}
\frac{d\beta_\pm}{dt}=-\frac{p_\pm e^{3\Omega}}{12\pi R_0^3}
\end{equation}
a stable isotropic state needs $p_\pm e^{3\Omega}\rightarrow 0$. We now look for the conditions allowing this limit.\\
Let us assume that isotropy leads to a static Universe, i.e. $\Omega$ tends toward a constant. Then $p_\pm$ must tend toward zero such that $p_\pm e^{3\Omega}$ vanishes. However, from (\ref{lapse}) and (\ref{ppp}-\ref{pmp}), it comes that $dp_\pm/dt\propto \partial V/\partial \beta_\pm e^{3\Omega}$. Hence for the Bianchi $I$, $VI_0$ and $VIII$ models, when $\beta_+$ and $\Omega$ tend toward some constants, the conjugate momentum diverges with the proper time $t$ since $\dot{p}_\pm$ tends toward a non vanishing constant and $p_+e^{3\Omega}\rightarrow \infty$. Thus from the reasonable assumption that isotropy happens when $t$ diverges, we deduce that isotropisation can not lead to a static Universe for these three models (it would be deeply unnatural that a static Universe ends for a finite value of $t$). For the $VII_0$ and $IX$ models, the demonstration is not so simple since then, when $\beta_\pm\rightarrow 0$ and $\Omega\rightarrow const$, $\partial V/\partial \beta_\pm\rightarrow 0$ and $\dot{p}_\pm$ vanishes. We will show below that for these models also, isotropy is not possible when $\Omega$ tends toward a constant. If now we assume that $\Omega$ diverges, nothing at this stage prevents the asymptotical vanishing of $p_\pm e^{3\Omega}$. Moreover, since $\beta_\pm$ tend toward some constants for a diverging value of $\Omega$, we deduce that $d\beta_\pm/d\Omega\rightarrow 0$ otherwise $\beta_\pm$ would diverge with $\Omega$. Consequently, isotropisation should arise when $\Omega\rightarrow \pm\infty$, $d\beta_\pm/d\Omega$ and $pe^{3\Omega}\rightarrow 0$. These conditions are independent each other and of the considered Bianchi class $A$ models. We will have to check if each of them is respected for each presumed isotropic equilibrium state.\\
Let us compare them with isotropisation conditions defined by Collins and Hawking \cite{ColHaw73}. First, as in this last paper, we have assumed that isotropisation arises when $t\rightarrow \infty$. Second, in the next sections, for each Bianchi models, we will show that isotropisation needs $\Omega\rightarrow -\infty$. This is the first condition that defines isotropisation in Collins and Hawking's paper and which implies that Universe expands indefinitely. Third, the fact that $d\beta_\pm/dt$ and $d\beta_\pm/d\Omega$ tend toward zero satisfies their third condition meaning that "the anisotropy in the locally measured Hubble constant tends to zero". Thus from the necessary but not sufficient conditions stating that isotropisation needs $d\beta_\pm/dt\rightarrow 0$ and considering the field equations, we recover two of the four conditions of the Collins and Hawking definition for isotropy. It shows the physical meaning of the limit $p_\pm e^{3\Omega}\rightarrow 0$ regarding isotropisation. Moreover, assuming that $\beta_\pm$ tend toward some constants means that the matrix $\beta$ whose components are the $\beta_{ij}$ becomes a constant $\beta_0$ which is the fourth condition defining isotropy in \cite{ColHaw73}.\\
In what follows, we study isotropic equilibrium states of each Bianchi class A model with curvature.
%------------------------------------------------------------------------------------------------------------------------------------------------------------------------------------------%
\subsection{The Bianchi type $II$ model} \label{s11}
%------------------------------------------------------------------------------------------------------------------------------------------------------------------------------------------%
\subsubsection{Field equations} \label{s111}
%------------------------------------------------------------------------------------------------------------------------------------------------------------------------------------------%
To study the equilibrium points corresponding to asymptotic isotropic states, we will use the following variables, specific to the Bianchi type $II$ model:
%-------------------------EQUATION-------------------------------------%
\begin{equation} \label{v1}
x_\pm=p_\pm H^{-1}
\end{equation}
%-------------------------EQUATION-------------------------------------%
\begin{equation} \label{v2}
y=\pi R_0^3 \sqrt{U}e^{-3\Omega}H^{-1}
\end{equation}
%-------------------------EQUATION-------------------------------------%
\begin{equation} \label{v3}
z=p_{\phi}\phi(3+2\omega)^{-1/2}H^{-1}
\end{equation}
%-------------------------EQUATION-------------------------------------%
\begin{equation} \label{v4}
w=\pi R_0^2 e^{-2\Omega+2(\beta_++\sqrt{3}\beta_-)}H^{-1}
\end{equation}
They are independent since $x_\pm$, $y$, $z$ and $w$ are respectively functions of the independent quantities $p_\pm$, $\phi$, $p_\phi$ and $\beta_\pm$. Then, the Hamiltonian (\ref{hamiltonien}) yields:
%-------------------------EQUATION-------------------------------------%
\begin{equation} \label{contrainte}
x_+^2+x_-^2+24y^2+12z^2+12w^2=1
\end{equation}
We will consider this last expression as a constraint. It shows that the 5 variables $(x_\pm,y,z,w)$ are normalised. They allow us to rewrite the field equations as a first order equations system in the following way:
%-------------------------EQUATION-------------------------------------%
\begin{equation} \label{eq1}
\dot{x}_+=72y^2x_++24w^2x_+-24w^2
\end{equation}
%-------------------------EQUATION-------------------------------------%
\begin{equation} \label{eq2}
\dot{x}_-=72y^2x_-+24w^2x_- -24\sqrt{3}w^2
\end{equation}
%-------------------------EQUATION-------------------------------------%
\begin{equation} \label{eq3}
\dot{y}=y(6\ell z+72y^2-3+24w^2)
\end{equation}
%-------------------------EQUATION-------------------------------------%
\begin{equation} \label{eq4}
\dot{z}=y^2(72z-12\ell)+24w^2z
\end{equation}
%-------------------------EQUATION-------------------------------------%
\begin{equation} \label{eq5}
\dot{w}=2w(x_++\sqrt{3}x_-+12w^2+36y^2-1)
\end{equation}
with $\ell=\phi U_\phi U^{-1}(3+2\omega)^{-1/2}$. Since we want to keep $\omega$ and $U$ undetermined, we will not explicit the form of $\ell(\phi)$. Then, the above system could not seem autonomous because $\ell=\ell(\phi)$ and thus it would be meaningless to look for its equilibrium point. However, it exists two possibilities such that it becomes autonomous. The first one is to consider $\ell$ as a function of $(x,y,z,w)$ rather than $\phi$. Such considerations are often used when one look for exact solutions of field equations: for instance, instead of considering the potential as a function of the scalar field, it can be easier to find exact solutions by assuming it depends on the metric functions\cite{Arb99}. However, in the general case, $\ell$ may not be written in this way in the whole range of $\phi$. The second possibility, which applies whatever $\ell$, is to consider an additional first order equation for $\phi$ which is derived from (\ref{phip}): $\dot{\phi}(z,\phi)$. Then, the system is autonomous. However, the constraint equation shows that scalar field equilibrium is not necessary for isotropisation contrary to other variables: $\dot\phi$ and $\phi$ can diverge whereas isotropy asymptotically occurs. Hence, we conclude that whatever the way we use such that the system be autonomous, equilibrium states are only determined by zeros of the system (\ref{eq1}-\ref{eq5}). The same reasoning may be applied for each Bianchi model. In what follows, we will use (\ref{eq1}-\ref{eq5}) to get the expressions for equilibrium points as some functions of $\ell(\phi)$ and then $\dot{\phi}(z,w)$ to get the asymptotical behaviour of the scalar field.
\subsubsection{Isotropic equilibrium states} \label{s112}
%------------------------------------------------------------------------------------------------------------------------------------------------------------------------------------------%
\emph{Equilibrium points}\\
\\
Several equilibrium points exist and we have to select those representing an isotropic equilibrium state. Immediately, we observe that the necessary conditions for isotropisation $d\beta_\pm/d\Omega\rightarrow 0$, imply that $x_\pm\rightarrow 0$ near equilibrium\footnote{Note that we have then, near equilibrium, $\dot{x}_\pm\rightarrow 0$ and $x_\pm\rightarrow 0$. This is a consequence of the field equations and values of the equilibrium points near isotropy. It means that $x_\pm$ should be integrable in the Lebesgue sense in the neighbourhood of equilibrium. We will see that it is actually the case when we will calculate the asymptotical behaviours of $x_\pm$.}. Since the equations (\ref{betapm}) and the definitions of the variables $x_\pm$ will be identical for all the Bianchi models, these limits will be the same for each of them. Thus, we will systematically discard any equilibrium point with $x_\pm\not =0$ or which is not defined by real values.\\
All the equilibrium points of the Bianchi type $II$ model are summarised in the appendix. The first one is defined by $(y,w)=(0,0)$. Let us show that it is not consistent with isotropy. From equations (\ref{ppp}-\ref{pmp}), we derive that $p_+-\sqrt{3}p_-=p_0$, $p_0$ being an integration constant. Then, using (\ref{betapm}), it comes that $\dot{\beta}_+-\sqrt{3}\dot{\beta}_-=p_0H^{-1}$. Thus, isotropisation needs a diverging Hamiltonian but for the zero measure case with $p_0=0$. If now we consider (\ref{eq5}) near $(y,w)=(0,0)$, we deduce that $w$ behaves as $e^{-2\Omega}$ and vanishes when $\Omega\rightarrow +\infty$. Introducing this expression in (\ref{v4}), we derive that $H$ should asymptotically behave as $e^{2(\beta_++\sqrt{3}\beta_-)}$ and thus should tend toward a constant value when Universe approaches isotropy. This contradict the fact that the Hamiltonian have to diverge and thus the set of points $(y,w)=(0,0)$ is not compatible with isotropisation.\\
The second set of points is not real and such that $x_\pm\not =0$. Hence, it does not correspond to an isotropic equilibrium state and we discard it. The third set of points is such that $x_\pm=0$ and respects the constraint. It writes $(x_{+},x_{-},y,z,w)=(0,0,\pm \sqrt{3-\ell^2}(6\sqrt{2})^{-1},\ell/6,0)$ and is real if $\ell^2<3$, implying that isotropisation is not possible if this last quantity diverges\footnote{We will not take into account the case for which $\ell^2$ would have a chaotic behaviour such that it stays smaller than $3$.}. Indeed since asymptotically $z$ behaves as $\ell/6$, it means that an isotropic stable state needs $\ell$ to tend to a constant value with no limit cycle otherwise $\dot{z}\not\rightarrow 0$ when $\Omega\rightarrow -\infty$. This is the only set of points representing an isotropic stable state and the only one we will consider below. The fourth set of points does not represent a stable isotropic state except if $\ell\rightarrow 1$. Then it tends toward the previous one. However, we will see below that the value $\ell=1$ does not allow for isotropisation. Thus we discard it.\\
\\
\emph{Monotonic functions}
\\
\\
We rewrite the equation (\ref{hamp}) with the normalised variables:
%--------------Equation-----------------------%
\begin{equation}\label{hamp1}
\dot{H}=-H(72y^2+24w^2)
\end{equation}
Hence the Hamiltonian is a monotonic function of $\Omega$ with a constant sign. Then, from the lapse function expression (\ref{lapse}), we deduce that $\Omega$ is a monotonic function of the proper time $t$. Therefore, if the Hamiltonian is initially positive (negative), $\Omega\rightarrow -\infty$ corresponds to late time(early time). We will not consider the case $\Omega\rightarrow +\infty$ since we will show below that it does not lead to isotropy. We conclude that late times isotropisation initially needs $H>0$: this is the only necessary initial condition for this behaviour. Moreover, the Hamiltonian being of constant sign, it is the same for the variables $y$ and $w$.\\
\\
\emph{Asymptotic behaviours}
\\
\\
We wish to evaluate the behaviours of some quantities in the neighbourhood of the equilibrium points $(x_{+},x_{-},y,z,w)=(0,0,\pm \sqrt{3-\ell^2}(6\sqrt{2})^{-1},\ell/6,0)$, i.e. when we approach isotropy in $\Omega\rightarrow -\infty$. Approximating (\ref{eq5}) near equilibrium by $w(1-\ell^2)$, we find that asymptotically $w$ behaves as $e^{(1-\ell^2)\Omega}$.

From this last expression and approximating (\ref{eq1}) by $(3-\ell^2)x_+-24w^2$, we deduce that $x_+$ behaves as the sum of two terms $e^{2(1-\ell^2)\Omega}$ and $e^{(3-\ell^2)\Omega}$. Since isotropy needs $x_+\rightarrow 0$ and $\ell^2<3$, we derive it only occurs if $\Omega\rightarrow -\infty$ and $\ell^2< 1$. These two limits are in accordance with the vanishing of $x_\pm$ and $w$ which is necessary to reach the equilibrium isotropic state. Consequently $x_\pm$ asymptotically behave as $e^{2(1-\ell^2)\Omega}$. Let us note that the two limits $x_\pm\rightarrow 0$ and $\Omega\rightarrow -\infty$, necessary for isotropisation, are compatible and justify our assumption that it takes place for a diverging value of $t$: when $\Omega\rightarrow -\infty$ and the Universe isotropises, it is expanding and there is no physical meaning to consider that it ends for a finite value of the proper time.\\
Let us show that the value $\ell^2=1$ has to be discarded. If $1-\ell^2\rightarrow 0$ faster than $\Omega^{-1}$, $w$ tends toward a non vanishing constant and is not compatible with isotropy. If $1-\ell^2\rightarrow 0$ slower than $\Omega^{-1}$, from (\ref{hamp}) we deduce that near equilibrium $H\rightarrow e^{-2\Omega}$. Then from (\ref{ppp}), it comes that $p_+\rightarrow e^{-2\Omega}$ and it follows from (\ref{v1}) that $x_+$ tends toward a non vanishing constant. Hence, the limit $\ell^2\rightarrow 1$ is not compatible with isotropy. The above reasoning concerning the case for which $1-\ell^2\rightarrow 0$ faster than $\Omega^{-1}$ will stay valid for any Bianchi model. However, when $1-\ell^2\rightarrow 0$ slower than $\Omega^{-1}$, it will be valid only for Bianchi type $II$, $VI_0$ and $VIII$ models. For Bianchi type $VII_0$ and $IX$ models, the situation is different because when $\beta_\pm\rightarrow 0$ during isotropisation, $p_\pm$ varies slower than $e^{-2\Omega}$ since $\partial V/\partial \beta_\pm e^{4\Omega}\rightarrow 0$. However, we will show below that near equilibrium, the value $\ell=1$ can also be excluded.\\
Now it is possible to show that $p_\pm e^{3\Omega}$ vanishes when $\Omega\rightarrow -\infty$. Writing $\dot{p}_\pm/H$ as a function of $x_\pm$ and $w$ and using their asymptotical behaviours, we calculate that $\dot{p}_\pm/p_\pm$ tends toward the constant $-(1+\ell^2)$. Consequently $p_\pm e^{3\Omega}\rightarrow e^{(2-\ell^2)\Omega}\rightarrow 0$ when $\Omega\rightarrow -\infty$ and the necessary but not sufficient condition for isotropisation is respected. Moreover, from (\ref{hamp1}), we calculate that asymptotically $\dot{H}H^{-1}$ tends toward $\ell^2-3$. Thus $H\rightarrow e^{(\ell^2-3)\Omega}$ and diverges since $\ell^2<1$: therefore, although determined independently, the asymptotical behaviours of $p_\pm$ and $H$ agreed with these of $x_\pm=p_\pm H^{-1}$.

Concerning the scalar field, we can find a differential equation whose solution asymptotically behaves in the same way as $\phi$ when $\Omega\rightarrow -\infty$ by expressing equation (\ref{phip}) with the normalised variables and considering its asymptotical limit near equilibrium. It comes:
%--------------Equation-----------------------%
\begin{equation}\label{phiAs}
\dot{\phi}=\frac{2\phi^2U_\phi}{(3+2\omega)U}
\end{equation}
This important equation allows us to get the asymptotical behaviour of $\phi$ near equilibrium and consequently that of $\ell$.

From the asymptotical behaviour of $H$ and the expression (\ref{lapse}) for the lapse function, it is possible to get the isotropic part of the metric, $e^{-\Omega}$, as a function of the proper time. If $\ell^2$ tends toward a non vanishing constant, then $e^{-\Omega}\rightarrow t^{\ell^{-2}}$. If $\ell^2$ tends to $0$ faster than $(-\Omega)^{-1}$, $e^{-\Omega}$ behaves like an exponential. Let us show that it is always the case. Equation (\ref{phiAs}) can be rewritten as $d\phi/d\Omega=2\ell^2U(U_\phi)^{-1}$ and then $U\propto e^{2\int\ell^2 d\Omega}$ when $\Omega\rightarrow -\infty$. Introducing this expression and the expression for $H$ into (\ref{v2}), we get $y\propto e^{-\ell^2\Omega+\int\ell^2 d\Omega}$. Thus, if $\ell^2$ vanishes slower than $(-\Omega)^{-1}$, $y$ diverges or tends to $0$ instead of a non vanishing constant and there is no equilibrium.

Hence, when an isotropic equilibrium state is reached with $\ell\rightarrow 0$, $\ell^2$ always vanishes faster than $(-\Omega)^{-1}$ and Universe always tends toward a De Sitter one. This proof for $\ell=0$ relies on the asymptotic form of $\phi$, $H$ and the definition of $y$. It will be valid for all the Bianchi models since we will see that these 3 quantities keep the same forms each time the same set of equilibrium points is considered.\\
From (\ref{v2}) and the asymptotical forms for $\Omega(t)$, we deduce that the potential behaves as $t^{-2}$ when $\ell$ tends toward a non vanishing constant, or as a non vanishing constant otherwise (the same behaviour held when we consider the Bianchi type $I$ model\cite{Fay01}). Concerning the 3-curvature which can be expressed as $R^{(3)}=e^{2\Omega+4(\beta_++\sqrt{3}\beta_-)}$, it is obvious that near isotropy, it tends to zero, showing that the Universe becomes spatially flat.\\
Let us note the importance of the potential for isotropisation of the Bianchi type $II$ model. If we consider $U=0$, we get from the field equations (\ref{ppp}) and (\ref{hamp}) that $H=p_++p_0$ and thus $\dot{\beta}_+=1-p_0H^{-1}$, $p_0$ being an integration constant. It follows that $\dot{\beta}_+$ does not asymptotically vanish and isotropy can not be reached.\\
\\
\emph{Partial equilibrium\footnote{I thank one of the referees criticisms from which this section is inspired.}}\\
\\
Above we have defined the isotropic equilibrium states such that all the variables reach equilibrium. However, this last statement is not mandatory: all of them have not to reach equilibrium such that $x_\pm\rightarrow 0$ in a stable way. In this case, as $\Omega\rightarrow -\infty$, the concerned variables would stay bounded but their derivatives would not asymptotically vanish: they should oscillate indefinitely (around a constant or not) and thus, their derivatives should also oscillate around zero without beeing damped. What are the variables able to behave in this way?\\
We can reasonably assume that it is not $x_\pm$, otherwise it would mean that the variation of the Hubble constant would be anisotropic. Since $x_\pm$ and $\dot x_\pm$ have to vanish, it implies that $w$ also asymptotically vanishes and consequently, $\dot w\rightarrow 0$. Finally the only variables whose equilibrium is not necessary to isotropy and whose derivatives could be oscillating are $y$ and $z$. Under these assumptions, what about $\ell$ behaviour?\\
Equation (\ref{eq4}) writes asymptotically $\dot z = y^2(72z-12\ell)$. Since we have assumed that $\dot z$ was oscillating, it shows that $\ell$ can not tend to a constant, diverge monotonically or diverge oscillatory if the oscillations are not large enought($\ell=\Omega^{1/3}+\cos \Omega$ as instance) otherwise the sign of $\dot z$ would be constant. Consequently, only oscillatory $\ell$ with sufficiently large amplitudes and not tending to a constant may allow oscillations of $z$ too ($\ell=\Omega \sin \Omega$ or $\ell=n \cos \Omega$ with $n$ a constant larger than the largest amplitude of $6z$ as instance). In this case, the results of the previous sections do not apply since all the variables do not reach equilibrium but an isotropic equilibrium state eventually occurs with $(x_\pm,w)$ only reaching equilibrium. Then, since $\ell$ can be a regular functions as well as having a chaotic behaviour, it is not possible to give more characteristics about it or the way $x_\pm$ would reach equilibrium. Hence, the main result of this subsection is mainly a limitation of the previous subsections results which will be valid for all the Bianchi class A models.\\
Let us examine the folowing example for the Bianchi type II model:\\
Since $x_\pm$ and $w$ vanish, asymptotically the constraint is $24y^2+12z^2=1$. In this limit, considering that (\ref{eq4}) is essentially equivalent to the constraint equation under (\ref{eq3}), (\ref{eq3}) is the only nontrivial equation. In this asymptotic reduction, (\ref{eq3}) can be regarded as  the equation for $\ell$ in terms of $y$, which can be written as:
$$
\ell=\frac{\dot v}{2+v}\frac{1}{\sqrt{1-v}}+\sqrt{1-v},
$$
where $v=2(36y^2-1)$ and $6z=\sqrt{1-v}$. For any function for $v(\Omega)$, the equation
$$
\dot \phi=\frac{12\phi}{\sqrt{3+2\omega}}z,
$$
determines $\phi$ as a function of $\Omega$ for a given $\omega$. Then, the definition of $\ell$,
$$
\ell=\frac{\phi}{\sqrt{3+2\omega}}\frac{U_\phi}{U},
$$
determines $U$ as a function of $\phi$, provided that $\phi-\Omega$ relation is invertible. Here, if $v$ is bounded by a positive constant from below and if $1-v$ is non-negative, one can easily check that $x_\pm$ and $w$ have required asymptotic behavior. Thus, the only possible constraint on $v(\Omega)$ is the invertibility of the $\phi-\Omega$ relation. For example, the choice:
$$
v=v_0+\frac{1}{\Omega}\sin^2\Omega^3
$$
satisfies this condition, if $v_0$ is a constant in the range $0<v_0<1$. However, for this choice, $\ell$, $\dot y$ and $\dot z$ diverge oscillatorily as $\Omega\rightarrow -\infty$, although the isotropization condition $x_\pm\rightarrow 0$ as $\Omega\rightarrow -\infty$ is satisfied and $(y,z)$ are bounded.
%------------------------------------------------------------------------------------------------------------------------------------------------------------------------------------------%
\subsection{The Bianchi type $VI_0$ and $VII_0$ models} \label{s12}
The results are similar to those of Bianchi type $II$ model.
%------------------------------------------------------------------------------------------------------------------------------------------------------------------------------------------%
\subsubsection{Field equations} \label{s121}
We will use the following variables for both Bianchi type $VI_0$ and $VII_0$ models:
\begin{eqnarray}
&x_\pm=p_\pm H^{-1}&\\
&y=\pi R_0^3e^{-3\Omega}U^{1/2}H^{-1}&\\
&z=p_\phi\phi (3+2\Omega)^{-1/2}H^{-1}&\\
&w_\pm=\pi R_0^2e^{-2\Omega+2\beta_{+}\pm 2\sqrt{3}\beta_-} H^{-1}\label{var41}&\\\nonumber
\end{eqnarray}
They are independent since they respectively depend on $p_\pm$, $\phi$, $p_\phi$ and a combining of $\beta_\pm$. The definitions of $x_\pm$, $y$ and $z$ are the same as these of the Bianchi type $II$ model. The Hamiltonian is written:
\begin{equation}
x_+^2+x_-^2+24y^2+12z^2+12(w_+\pm w_-)^2=1\label{contrainte1}
\end{equation}
and the field equations become:
\begin{eqnarray}
&\dot{x}_+=72y^2x_++24(x_+-1)(w_-\pm w_+)^2\label{eq12}&\\
&\dot{x}_-=72y^2x_-+24x_-(w_-\pm w_+)^2+24\sqrt{3}(w_-^2-w_+^2)\label{eq22}&\\
&\dot{y}=y(6\ell z+72y^2-3+24(w_-\pm w_+)^2)&\\
&\dot{z}=y^2(72z-12\ell)+24z(w_-\pm w_+)^2&\\
&\dot{w}_+=2w_+\left[x_++\sqrt{3}x_-+12(w_-\pm w_+)^2+36y^2-1\right]\label{eq52}&\\
&\dot{w}_-=2w_-\left[x_+-\sqrt{3}x_-+12(w_-\pm w_+)^2+36y^2-1\right]\label{eq62}&\\\nonumber
\end{eqnarray}
The $\pm$ symbols in equations (\ref{contrainte1}-\ref{eq62}) correspond respectively to the Bianchi type $VI_0$ and $VII_0$ models. For the first model, the constraint shows that the variables are normalised since $w_+$ and $w_-$ are positive. For the second one, because of the minus sign, both $w_+$ and $w_-$ could diverge. Then the constraint will be respected only if the sum $w_+-w_-$ tends toward a constant. We will show below that isotropy is only compatible with finite values of $w_\pm$. Consequently, the isotropic states we are looking for are reached for some bounded values of the variables whatever the Bianchi type $VI_0$ or $VII_0$ models. Now, we can also show that isotropisation of Bianchi type $VI_0$ and $VII_0$ models may not arise for a finite value of $\Omega$. Indeed, if $\Omega$ tends toward a constant when the proper time diverges, $d\Omega/dt$ vanishes and thus, from (\ref{lapse}), it comes that $H$ vanishes. But then, $w_\pm$ should diverge which is not compatible with the equilibrium as shown above. Hence, for the Bianchi type $VI_0$ and $VII_0$ models, isotropisation does not lead to a static Universe.
%------------------------------------------------------------------------------------------------------------------------------------------------------------------------------------------%
\subsubsection{Isotropic equilibrium states} \label{s122}
\emph{Equilibrium points}\\
\\
All the equilibrium points with finite values of $w_+$ and $w_-$ are referenced in the appendix. Following the same reasoning as for Bianchi type $II$ model, the only set of equilibrium points compatible with an isotropic stable state is $(x_{+},x_{-},y,z,w_+,w_-)=(0,0,\pm \sqrt{3-\ell^2}(6\sqrt{2})^{-1},\ell/6,0,0)$, implying that $\ell^2<3$. It is equivalent to the points found for Bianchi type $II$ model and, for the same reasons, $\ell$ have to tend to a constant with no limit cycle. Another interesting set of points is given by $(x_{+},x_{-},y,z,w_\pm,w_\pm)=((\ell^2-1)(\ell^2+8)^{-1},\pm\sqrt{3}(\ell^2-1)(\ell^2+8)^{-1},\pm \sqrt{12-3\ell^2}\left[2(\ell^2+8)\right]^{-1},3\ell(2\ell^2+16)^{-1},0,\pm\sqrt{-\ell^4+5\ell^2-4}\left[2(\ell^2+8)\right]^{-1})$. However, in this case $x_\pm\rightarrow 0$ only if $\ell^2\rightarrow 1$ and we recover the values of the previous set of points for this particular limit of $\ell^2$ which does not allow isotropisation as it will be shown below. Other sets exist but are complex valued and can thus be discarded. At last, as for Bianchi type $II$ model, let us show that the set of equilibrium points such that $(y,w_+,w_-)=(0,0,0)$ implies that $x_\pm$ do not vanish.\\
In this case we deduce from (\ref{eq52}-\ref{eq62}) that $w_\pm$ behave as $e^{-2\Omega}$ and thus isotropy would arise when $\Omega\rightarrow +\infty$. Then, using the definition (\ref{var41}) for $w_\pm$, we derive that $H$ should be a constant near isotropy and, considering equations (\ref{ppp}-\ref{pmp}), we find that asymptotically $\dot{p}_\pm$ should vary as $p_{0\pm}e^{-4\Omega}$ ($p_{0\pm}$ being some constants) for Bianchi type $VI_0$ or even slower for the $VII_0$ type since $\beta_\pm$ might tend toward some vanishing constants. It would follow that $p_\pm\rightarrow p_{1\pm}$, $p_{1\pm}$ being some integration constants. However, in this case $\dot{\beta}_\pm$ would tend asymptotically toward the constants $p_{1\pm}H^{-1}$ and isotropisation could not occur for a diverging value of $\Omega$. Hence, $(y,w_+,w_-)=(0,0,0)$ is not compatible with isotropisation except for the special case of zero measure $p_{1\pm}=0$.\\
\\
\emph{Monotonic functions}
\\
\\
What about monotonic functions? We can rewrite equation (\ref{hamp}) as follows:
%--------------Equation-----------------------%
\begin{equation}\label{hamp2}
\dot{H}=-H\left[72y^2+24(w_+\pm w_-)^2\right]
\end{equation}
As for the Bianchi type $II$ model, (\ref{hamp2}) shows that the Hamiltonian is a monotonic function of constant sign which determines if isotropisation occurs at early or late times depending if the Hamiltonian is initially negative or positive. Moreover, it follows that $y$ and $w_\pm$ are of constant sign.\\
\\
\emph{Asymptotic behaviours}\\
\\
Making the same approximation as for subsection \ref{s112}, we find that near an isotropic equilibrium state, $w_\pm\rightarrow e^{(1-\ell^2)\Omega}$. Then, assuming that isotropisation arises for $\Omega\rightarrow -\infty$ as we will show it below, it follows from (\ref{hamp2}) that $H\rightarrow e^{-2\Omega}$ when $\ell\rightarrow 1$ and then, near equilibrium, $w_\pm$ should tend toward some non vanishing constants. Thus the value $\ell=1$ does not agree with isotropisation.\\
Concerning $x_\pm$, from (\ref{eq12}-\ref{eq22}), we deduce that they asymptotically behave as the sum of two exponentials $e^{2(1-\ell^2)\Omega}$ and $e^{(3-\ell^2)\Omega}$, showing again that these quantities will tend toward zero only if $\Omega\rightarrow -\infty$ and $\ell^2<1$. These two limits allow vanishing of $w_\pm$, which is necessary to reach the equilibrium states.\\
All these elements show that near isotropy $w_\pm$ is bounded. Indeed, for $(x,y,z)$ to reach equilibrium, one only needs that $w_+-w_-\rightarrow 0$ whatever the particular asymptotical behaviours of $w_\pm$. Then only using this last limit, we have recovered the asymptotical behaviours for $x_\pm$ which implies that $\ell^2<1$. From (\ref{eq52}-\ref{eq62}) and these last expressions and condition, it is then possible to get the particular behaviours of $w_\pm$, which show that, near isotropy, these variables always vanish and are then bounded.\\
Again, we are able to calculate that $p_\pm e^{3\Omega}$ tends to $0$ as $e^{(2-\ell^2)\Omega}$, thus filling the necessary but not sufficient conditions for isotropisation we defined above. From equation (\ref{hamp2}) we derive that asymptotically $H$ behaves as $e^{(\ell^2-3)\Omega}$. The form of equation (\ref{phip}) for $\dot{\phi}$ being unchanged whatever the Bianchi model as well as those of $\ell$ and $z$ near equilibrium, we recover the same differential equation as (\ref{phiAs}) giving the asymptotical behaviour for the scalar field.\\
Since $H$ and $\dot{\phi}$ when $\Omega\rightarrow -\infty$, and $N$ and $y$ have the same forms as for the Bianchi type $II$ model, we find the same behaviours for $e^{-\Omega}$ and $U$ as a function of the proper time depending if $\ell$ tends or not toward a vanishing constant. Again, the 3-curvature $^{3}R$ tends to zero since $\beta_\pm$ become constant when $\Omega$ diverges negatively.
%------------------------------------------------------------------------------------------------------------------------------------------------------------------------------------------%
\subsection{The Bianchi type $VIII$ and $IX$ models} \label{s13}
%------------------------------------------------------------------------------------------------------------------------------------------------------------------------------------------%
\subsubsection{Field equations} \label{s131}
%------------------------------------------------------------------------------------------------------------------------------------------------------------------------------------------%
We will use the following variables:
\begin{eqnarray*}
&x_\pm=p_\pm H^{-1}&\\
&y=\pi R_0^3 e^{-3\Omega}U^{1/2}H^{-1}&\\
&z=p_\phi\phi (3+2\Omega)^{-1/2}H^{-1}&\\
&w_{p}=\pi R_0^2 e^{-2\Omega+2\beta_{+}}H^{-1}&\\
&w_{m}=\pi R_0^2 e^{-2\Omega-2\beta_{+}}H^{-1}&\\
&w_{-}=e^{2\sqrt{3}\beta_{-}}&\\
\end{eqnarray*}
The variables $x_\pm$, $y$ and $z$ are the same as those defined for the Bianchi type $II$ model. $w_p$ and $w_m$ are not independent because both are related to $\beta_+$. Near isotropy, we will have $w_m\propto w_p\propto e^{-2\Omega}H^{-1}$. $w_-$ is positive. The constraint equation is written:
\begin{eqnarray*}
&x_+^2+x_-^2+24y^2+12z^2+12\mbox[w_p^3(1+w_-^4)\pm 2 w_-(w_mw_p)^{3/2}(1+w_-^2)+&\\
&w_-^2(w_m^3-2w_p^3)\mbox](w_-^2w_p)^{-1}=1&\\
\end{eqnarray*}
and for the field equations it comes:
\begin{eqnarray}
&\dot{x}_+=72y^2x_++24\{w_p^3(x_+-1)(1+w_-^4)\pm w_-(1+2x_+)(w_mw_p)^{3/2}(1+w_-^2)&\nonumber\\
&+w_-^2\left[(2+x_+)w_m^3-2(x_+-1)w_p^3\right]\}(w_-^2w_p)^{-1}\label{eq13}&\\
&\dot{x}_-=72y^2x_-+24\{w_p^3\left[w_-^4(x_--\sqrt{3})+x_-+\sqrt{3})\right]\pm w_-(w_mw_p)^{3/2}\mbox{[}w_-^2&\nonumber\\
&(-\sqrt{3}+2x_-)+(\sqrt{3}+2x_-)\mbox{]}+w_-^2x_-(w_m^3-2w_p^3)\}(w_-^2w_p)^{-1}\label{eq23}&\\
&\dot{y}=y\{6\ell z+72y^2-3+24\mbox{[}w_p^3(1+w_-^4)\pm 2(w_mw_p)^{3/2}w_-(1+w_-^2)+&\nonumber\\
&w_-^2(w_m^3-2w_p^3)\mbox{]}(w_-^2w_p)^{-1}\}&\\
&\dot{z}=y^2(72z-12\ell)+24z\mbox{[}w_p^3(1+w_-^4)\pm 2(w_mw_p)^{3/2}w_-(1+w_-^2)+&\nonumber\\
&w_-^2(w_m^3-2w_p^3)\mbox{]}(w_-^2w_p)^{-1}&\label{eq43}\\
&\dot{w}_p=w_p\{-2+2x_++72y^2+24\mbox{[}w_p^3(1+w_-^4)\pm2w_-(w_mw_p)^{3/2}(1+w_-^2)&\nonumber\\
&+w_-^2(w_m^3-2w_p^3)\mbox{]}(w_-^{2}w_p)^{-1}\}&\label{eq53}\\
&\dot{w}_m=w_m\{-2-2x_++72y^2+24\mbox{[}w_p^3(1+w_-^4)\pm2w_-(w_mw_p)^{3/2}(1+w_-^2)&\nonumber\\
&+w_-^2(w_m^3-2w_p^3)\mbox{]}(w_-^{2}w_p)^{-1}\}&\label{eq63}\\
&\dot{w}_-=2\sqrt{3}w_-x_-&\label{eq73}\\\nonumber
\end{eqnarray}
$\pm$ being related respectively to the Bianchi type $VIII$ and $IX$ models. For sake of completeness, we have written differential equations for each variable $w_p$ and $w_m$. However, they are equivalent. The constraint shows that the variables are not necessarily normalised: if one of them diverges near isotropy, it have to be counterbalanced by the divergence of $w_p$ and $w_m$. Thus, if we show that isotropy does not arise for unbounded values of $w_p$ and $w_m$, it will mean that it only happens for some finite values of the variables.\\
To reach this goal, we will write that $w_p\rightarrow w_m\rightarrow w$ and $w_-\rightarrow 1$. This is justified because isotropy needs $\beta_\pm\rightarrow 0$ and we will see below that an isotropic equilibrium state effectively implies $w_-\rightarrow 1$. In this case, for Bianchi type $VIII$ model, all the variables in the constraint are positives and thus bounded. Concerning the Bianchi type $IX$ model, let us assume that $w$ diverges. Then, putting $x_\pm=0$, from the constraint we have asymptotically $3w^2\rightarrow 2y^2+z^2-1/12$ and from the equation for $\dot w$,
$3w^2\rightarrow 3y^2-1/12$, implying that asymptotically $z^2\rightarrow y^2$ and diverges as $w^2$. However, with these limits we get from the equations for $\dot y$ and $\dot z$ that $\dot y\rightarrow 6\ell z^2-3z$ and $\dot z\rightarrow -12\ell z^2+2z$. Then equilibrium for $y$ and $z$ can only be reached if $z\rightarrow 0$ which is in disagreement with the assumption on the divergence of $w$. Hence, an isotropic equilibrium state is not possible if $w_p$ and $w_m$ diverge. It follows, in the same way as for the Bianchi type $VI_0$ and $VII_0$ models, that isotropisation of Bianchi type $VIII$ and $IX$ models for a finite value of $\Omega$ is impossible.\\
We can also show that $w_p$ and $w_m$ may not tend toward some non vanishing constants. Let us assume that it is actually the case and consider 2 constants $w$ and $\alpha$ such that $w_p\rightarrow w$ and $w_m\rightarrow \alpha w$. We introduce these limits in equations (\ref{eq13}-\ref{eq23}) with $x_\pm=0$ and get respectively:
\begin{eqnarray}
&\dot{x}_+=-24w^2(1+w_-\alpha^{3/2}(1+w_-^2)-2w_-^2(1+\alpha^3)+w_-^4)w_-^{-2}&\\
&\dot{x}_-=-24\sqrt{3}w^2(w_-^2-1)(1-\alpha^{3/2}w_-+w_-^2)w_-^{-2}&\\\nonumber
\end{eqnarray}
Then, for Bianchi type $VIII$ model, we derive that equilibrium for $x_\pm$ will be reached only if $\alpha$ tends toward a complex value $(-1)^{2/3}$ or/and $w_-$ is negative, which is impossible. For the Bianchi type $IX$ model, equilibrium for $x_\pm$ may be reached if $w_p\rightarrow w_m$ (i.e. $\beta_\pm\rightarrow 0$) and $w_-\rightarrow 1$. Then, looking for the equilibrium points, the only ones which may be real and such that $w_p$ and $w_m$ be different from $0$ are $(x_+,x_-,y,z,w_p,w_m,w_-)=(0,0,\pm(6\ell)^{-1},(6\ell)^{-1},\pm (1-\ell^2)^{1/2}(6\ell)^{-1})$. They check the constraint equation and are real if $\ell^2<1$. Then we calculate that $w_p$ and $w_m$ behave like $\pm\left[(1-\ell^2)(1-e^{\frac{4\Omega(\ell^2-1)+\omega_0}{\ell^2}}+36\ell^2)^{-1}\right]^{1/2}$ and thus reach equilibrium when $\Omega\rightarrow +\infty$. Meantime, starting from this last expression and introducing it in the equation for $x_+$, it comes that $x_+$ tends toward a complex value when $\Omega\rightarrow +\infty$ and thus these equilibrium points are excluded. Numerical simulations seem to confirm the absence of equilibrium for these values of $(x_+,x_-,y,z,w_p,w_m,w_-)$.\\
%------------------------------------------------------------------------------------------------------------------------------------------------------------------------------------------%
\subsubsection{Isotropic equilibrium states} \label{s132}
\emph{Equilibrium points}\\
\\
To find the equilibrium points we have to consider the equations (\ref{eq13}-\ref{eq43}), (\ref{eq73}) and one of the equations (\ref{eq53}) or (\ref{eq63}) since both $w_m$ and $w_p$ depend on  $\beta_+$. However the solutions of the equations system thus defined can not be easily calculated. Consequently, we will only take into account the solutions such that $(x_\pm,w_p,w_m)=(0,0,0)$ and which are compatible with isotropy. Then the solutions reduce to the set of equilibrium points $(x_{+},x_{-},y,z,w_p,w_m,w_-)=(0,0,\pm \sqrt{3-\ell^2}(6\sqrt{2})^{-1},\ell/6,0,0,1)$ which will be real if $\ell^2<3$ and respect the constraint equation. It is equivalent to the sets found for the previous models and once again, $\ell$ have to tend to a constant with no limit cycle such that equilibrium be reached. Note that it is such that $\beta_- \rightarrow 0$ since $w_-\rightarrow 1$.\\
\\
\emph{Monotonic functions}\\
\\
We can rewrite (\ref{hamp}) in the following form:
%-----------------------EQUATION-------------------------------------%
\begin{eqnarray}\label{hamp3}
&\dot{H}=-H\mbox{[}72y^2+24(\pm 2\frac{w_p^{1/2}w_m^{3/2}}{w_-}\pm 2w_p^{1/2}w_m^{3/2}w_--2w_p^2+\frac{w_p^2}{w_-^2}+&\nonumber\\
&w_p^2w_-^2+\frac{w_m^3}{w_p})\mbox{]}&\\
\end{eqnarray}
We immediately see that it is not a monotonic function and that its sign is indefinite. Thus $\Omega$ is not a monotonic function of $t$ and it is not possible to determine if isotropisation, corresponding to $\Omega\rightarrow -\infty$, arises at early or late proper times.\\
\\
\emph{Asymptotic behaviours}\\
\\
Near equilibrium, it is possible to approximate equation (\ref{eq53}) by $w_p(1-\ell^2)$ implying that $w_p$ tends toward $e^{(1-\ell^2)\Omega}$. The same conclusion arises for $w_m$. In the same way as the previous subsection, one can show that the value $\ell^2=1$ is not agreed with isotropisation. Introducing asymptotical expressions for $w_p$ and $w_m$ in the equations (\ref{eq13}-\ref{eq23}), we find that $x_\pm$ behave as the sum of two exponentials, $e^{2(1-\ell^2)\Omega}$ and $e^{(3-\ell^2)\Omega}$. Thus, once again, isotropisation needs $\Omega\rightarrow -\infty$ and $\ell^2<1$ implying that $x_\pm$ behave as $e^{2(1-\ell^2)\Omega}$. As for Bianchi type $II$ model, it is possible to show that $p_\pm e^{3\Omega}\rightarrow e^{(2-\ell^2)\Omega}$ and thus vanish. Moreover $H$ behaves again as $e^{(\ell^2-3)\Omega}$. The Hamiltonian equation (\ref{phip}) for $\dot{\phi}$ being independent of considered Bianchi model and the definition and asymptotical value of $z$ being the same as for Bianchi type $II$ model, we find the same differential equation (\ref{phiAs}), giving asymptotically the behaviour of the scalar field. Anew, since $H$ and $\dot{\phi}$ when $\Omega\rightarrow -\infty$, and $N$ and $y$ have the same forms as for Bianchi type $II$ model, the discussion about the forms of the metric functions  near equilibrium is the same and they behave as power or exponential law of the proper time depending on the asymptotical value of $\ell$.
%------------------------------------------------------------------------------------------------------------------------------------------------------------------------------------------%
\section{Discussion} \label{s3}
We have found some necessary conditions for isotropisation of Bianchi class $A$ models with curvature for a minimally coupled scalar tensor theory. We have seen that the Universe has to expand ($\Omega\rightarrow -\infty$), justifying the assumption that $t$ should be diverging, and that the ratio between the conjugate momentum and the Hamiltonian should vanish. Our results do not concern the class of theories for which $\ell$ prevents the equilibrium of $z$ and $y$. As shown in subsection \ref{s112}, such $\ell$ should be oscillating with significant amplitude and not tending to a constant. Hence, they concern the $\ell$ which tend to a constant, diverge monotonically or even with negligible oscillations. In these cases, our main result is:\\
\\
\emph{A necessary condition for isotropisation of Bianchi class $A$ models with curvature for General Relativity plus a massive scalar field, whatever the considered Brans-Dicke coupling function and potential, will be that $\phi U_\phi U^{-1} (3+2\omega)^{-1/2}$ tends to a constant $\ell$ such that $\ell^2<1$. For Bianchi type $II$, $VI_0$ and $VII_0$ models, it arises at late times if the Hamiltonian is positive, at early times otherwise. For Bianchi type $VIII$ and $IX$ models, the time of isotropisation is undetermined. If isotropisation arises with $\ell\not =0$ the metric functions tend toward a power inflationary law $t^{\ell^{-2}}$ and the potential vanishes as $t^{-2}$. If it arises as $\ell=0$, the Universe tends toward a De Sitter model and the potential to a constant. In any case, isotropisation requires late time accelerated expansion and the Universe becomes spatially flat.
}\\
\\
Necessary condition for isotropisation determines an asymptotical limit that the scalar field have to respect. It can be compared to the limit required such that scalar tensor theories be compatible with solar system tests when $U=0$, i.e. $\omega\rightarrow \infty$ and $\omega_\phi\omega^3\rightarrow 0$. To evaluate $\ell$, we need to know the asymptotical behaviour of the scalar field. It comes:\\
\\
\emph{The value of the scalar field when the Universe reaches an equilibrium isotropic state with an asymptotically constant $\ell$ is the value of the function $\phi$ defined by $\dot{\phi}=2\phi^2U_\phi(3+2\omega)^{-1}U^{-1}$ when $\Omega\rightarrow -\infty$.\\}
\\
Although Bianchi type $IX$ model contains the closed FLRW solutions, when the Universe isotropises and we consider a minimally coupled and massive scalar field, it is infinitely expanding. Moreover, the common late time attractor of all the isotropising solutions is not oscillating. This fact may seem astonishing for the Bianchi type $VIII$ and $IX$ models. However, in \cite{BarGas01}, it has been observed that despite the mixmaster behaviour of Bianchi type $VIII$ model at early time, its late time behaviour can be non oscillating. If we compare the results got when no curvature is present \cite{Fay01} with those of this paper, few differences appear. The asymptotical behaviours of the scalar field and isotropic part of the metric are the same in both papers, partly because the Hamiltonian and lapse function asymptotically behave in the same way. However the variations of the functions describing the anisotropy, $\beta_\pm$, are different since the conjugate momenta are not constant in presence of curvature. The fundamental difference comes from the interval of $\ell$ allowing for isotropy. For the Bianchi type $I$ model, it was $\ell^2<3$ and decelerated  dynamics was possible. For the models with curvature, we have $\ell^2<1$, implying that isotropisation requires late time accelerated expansion. This is due to the presence of curvature which reduces the interval of values of $\ell$ related to the Bianchi type $I$ model. Hence, late time accelerated expansion finds a natural explanation through the fact that our Universe is isotropic. Other problems are naturally solved by isotropisation: asymptotically, the 3-curvature vanishes thus solving the flatness problem. It comes from the fact that during isotropisation $\beta_\pm$ tend toward a constant whereas $\Omega\rightarrow -\infty$. In the same way, the small value of the cosmological constant could be explained by the fact that when Universe isotropises and $\ell$ does not vanish, the potential, which mimics a dynamical cosmological constant vanishes. If $\ell$ vanishes, the potential tends to a non vanishing constant but it is not necessary small except if we fine tune it.\\

To complete this study, let us write few words about Bianchi class $B$ models. Their Hamiltonian formulation is different from those of class $A$ and has been studied in \cite{RyaWal84}. It needs to redefine the divergence theorem in a non-coordinated basis. Then the Bianchi type $V$ Hamiltonian writes as the one of Bianchi type $I$ with an additional constraint $p_+=0$. Consequently for Bianchi type $V$ model which contains the  solutions of open FLRW model, isotropisation follows the same rules as those of Bianchi type $I$ model described in \cite{Fay01}. The nature of the other class $B$ Hamiltonians is totally different and will not be considered here.\\
\\
Let us examine some results usually considered in the literature.

The "No Hair theorem" of Wald\cite{Wal83} states that General Relativity with a scalar field and a cosmological constant isotropises toward a De Sitter Universe. Here, as for the Bianchi type $I$ model, when $\ell\rightarrow 0$ and if the minimally coupled scalar tensor theory isotropises\footnote{Do not forget that $\ell$ tending toward a constant is a necessary but not sufficient condition for isotropisation.}, it will tend toward a De Sitter Universe. This generalise the "No Hair theorem" which takes into account only the case $U=cte$ for which $\ell=0$.

It is really shocking that it exists only one set of equilibrium points shared by all the Bianchi models and representing the only possible isotropic stable equilibrium state. However, despite a careful analysis we have not found any additional points with such properties. One way to check if this statement is true is to select some special forms of $U$ and $\omega$ and then to verify if the conditions for isotropisation of the theory thus defined and the asymptotical value of $e^{-\Omega}$ are in agreement with our results. It can be easily done with the theory defined by an exponential potential $U=e^{k\phi}$ and a Brans-Dicke coupling function $\sqrt{3+2\omega}\phi^{-1}=\sqrt{2}$ whose isotropisation has been extensively studied in the literature using different methods. In this case $\ell^2=k^2/2$. We have collected the conclusions of different papers and have compared them with ours. In \cite{ColIbaHoo97, KitMae92}, it is shown that isotropisation arises at late time when $k^2<2$ (except the contracting Bianchi type $IX$ models) and lead to a De Sitter Universe when $k=0$ or to a power law of the form $t^{2k^{-2}}$ for the metric functions otherwise. If $k^2>2$, the Bianchi type $I$, $V$, $VII$ and $IX$ models might isotropise at late times. Concerning the Bianchi type $I$ model, we have shown in \cite{Fay01} that a necessary condition for isotropisation will be $k^2<6$ but it was impossible for larger values. For the models of class $A$ with curvature, from the present paper we deduce that isotropisation is possible only when $k^2<2$ and always comes with late time accelerated expansion. The asymptotic behaviour of the metric functions is in accordance with that of \cite{ColIbaHoo97, KitMae92}. A difference is that Bianchi type $VII_0$ and $IX$ should not isotropise if $k^2>2$. Concerning the Bianchi type $VI_0$ model, our results agree with these of \cite{ChiLab98}. For the Bianchi type $V$ model, they are the same as these of the Bianchi type $I$ model in accordance with \cite{ColIbaHoo97}. Hence, concerning the special case of an exponential potential, there are few differences between our results and those of others papers. It seems to confirm the presence of a unique set of equilibrium points shared by all the Bianchi class $A$ models and representing an isotropic equilibrium state. Of course, the case of an exponential potential could be a particular one and thus other types of potentials should be studied to check the results of the present paper. Note that, isotropic state is not the only possible late time equilibrium state. As written above or shown in the appendix, other ones exist, for instance with $x_\pm\not = 0$, but they do not correspond to an isotropic Universe.\\

To conclude, Universe isotropisation requires late time accelerated expansion because of the curvature. Then, it becomes flat and the potential vanishes as $t^{-2}$ or tends toward a constant. These features fit well with the current observations and leave the door open to geometrical and physical generalisations of standard cosmological framework. In a next work, we will take into account the presence of a perfect fluid.
%------------------------------------------------------------------------------------------------------------------------------------------------------------------------------------------%
\ack
I thanks Mr Jean-Pierre Luminet for useful discussion and carefull reading of the manuscript. I also thanks anonymous referees for improving the manuscript.
%------------------------------------------------------------------------------------------------------------------------------------------------------------------------------------------%
\section{Appendix} \label{a1}
In this appendix, we present all the equilibrium points of Bianchi type $II$, $VI_0$ and $VII_0$ models.\\
\\
Bianchi type $II$ model :
\begin{itemize}
\item $(y,w)=(0,0)$
\item $(x_+,x_1,y,z,w)=(1,\sqrt{3},0,0,\pm i/2)$
\item $(x_+,x_1,y,z,w)=(0,0,\frac{\pm\sqrt{3-\ell^2}}{(6\sqrt{2})},\ell/6,0)$
\item $(x_+,x_1,y,z,w)=(\frac{\ell^2-1}{\ell^2+8},\sqrt{3}\frac{\ell^2-1}{\ell^2+8},\pm\frac{\sqrt{12-3\ell^2}}{2(\ell^2+8)},\frac{3\ell}{2(\ell^2+8)},\pm\frac{\sqrt{(\ell^2-1)(\ell^2-4)}}{2(\ell^2+8)})$
\end{itemize}
Bianchi type $VI_0$ and $VII_0$ models :
\begin{itemize}
\item $(y,w_+,w_-)=(0,0,0)$
\item $(x_+,x_-,y,w_+,w_-)=(1,0,0,w_+,w_+)$
\item $(x_+,x_-,y,z,w_+,w_-)=(1,-\sqrt{3},0,0,0,\pm i/2)$
\item $(x_+,x_-,y,z,w_+,w_-)=(1,\sqrt{3},0,0,\pm i/2,0)$
\item $(x_+,x_-,y,z,w_+,w_-)=(0,0,\frac{\pm\sqrt{3-\ell^2}}{(6\sqrt{2})},\ell/6,0,0)$
\item $(x_+,x_-,y,z,w_+,w_-)=(\frac{\ell^2-1}{\ell^2+8},-\sqrt{3}\frac{\ell^2-1}{\ell^2+8},\pm\frac{\sqrt{12-3\ell^2}}{2(\ell^2+8)},\frac{3\ell}{2(\ell^2+8)},0,\pm\frac{\sqrt{(\ell^2-1)(\ell^2-4)}}{2(\ell^2+8)})$
\item $(x_+,x_-,y,z,w_+,w_-)=(\frac{\ell^2-1}{\ell^2+8},\sqrt{3}\frac{\ell^2-1}{\ell^2+8},\pm\frac{\sqrt{12-3\ell^2}}{2(\ell^2+8)},\frac{3\ell}{2(\ell^2+8)},\pm\frac{\sqrt{(\ell^2-1)(\ell^2-4)}}{2(\ell^2+8)},0)$
\end{itemize}
%------------------------------------------------------------------------------------------------------------------------------------------------------------------------------------------%

\begin{table}[b] \label{tab}
\caption{Form of $V(\Omega,\beta_+,\beta_-)$ for each Bianchi model.}
\begin{indented}
\item[]\begin{tabular}{@{}ll}
\br
Bianchi type&$V(\Omega,\beta_+,\beta_-)$\\
$II$&$12\pi^2R_0^4e^{4(-\Omega+\beta_++\sqrt{3}\beta_-)}$\\
$VI_0$, $VII_0$&$24\pi^2R_0^4e^{-4\Omega+4\beta_+}(\cosh{4\sqrt{3}\beta_-}\pm 1)$\\
$VIII$, $IX$&$24\pi^2R_0^4e^{-4\Omega}\mbox{[}e^{4\beta_+}(\cosh{4\sqrt{3}\beta_-}-1)+$\\
&$1/2e^{-8\beta_+}\pm2e^{-2\beta_+}\cosh{2\sqrt{3}\beta_-}\mbox{]}$\\
\br
\end{tabular}
\end{indented}
\end{table}
\end{document}